\documentstyle[aps,graphicx,epsf,latexsym]{revtex} 
\makeatletter
\def\thepage{\@arabic\c@page}
\def\@pnumwidth{2em}
\def\REVTeX{REV\TeX}
\makeatother
\begin{document}
\twocolumn

\setlength{\unitlength}{1cm}

\title{Extended corresponding-states behavior for particles with variable range
attractions}
\author{Massimo G. Noro and Daan Frenkel}
\address{
FOM Institute for Atomic and Molecular Physics\\
Kruislaan 407, 1098 SJ Amsterdam, The Netherlands}
\maketitle

\makeatletter
\global\@specialpagefalse
\def\@oddhead{Typeset with \REVTeX{} \hfill Preprint}
\let\@evenhead\@oddhead
% run page numbers with copyright for first page
\def\@oddfoot{\reset@font\rm\hfill \thepage\hfill
\ifnum\c@page=1
  \llap{\protect\copyright{}
  Submitted to The Journal of Chemical Physics}%
\fi
} \let\@evenfoot\@oddfoot
\makeatother

\begin{abstract}
We propose an extension of the law of corresponding states that can be
applied to systems - such as colloidal suspensions - that have widely
different ranges of attractive interactions. We argue that, for such
systems, the ``reduced'' second virial coefficient is a convenient parameter
to quantify the effective range of attraction. This procedure allows us to
give a simple definition of the effective range of attraction of potentials
with different functional forms. The advantage of the present
approach is that it allows us to estimate  the relative
location of the liquid-vapor and solid-fluid coexistence curves
exclusively on basis of the knowledge of the pair-potential.
\end{abstract}

\vskip 1cm Van der Waals's Law of Corresponding States expresses the
fact that there are basic similarities in  the thermodynamic
properties of all simple gases.  Its essential feature is that if we
scale the thermodynamic variables that describe an equation of state
(temperature, pressure and volume) with respect to their values at the
critical point, all simple fluids obey the same  reduced equation of
state. Pitzer \cite{Pitzer.1} has given a molecular interpretation of
the Law of Corresponding States for classical monoatomic systems using
statistical mechanical arguments. This proof is restricted to
systems for which the total intermolecular potential can be written as
a sum over pair potentials in the form: 
\begin{equation}
U=\sum_{i,j}\epsilon \text{ }v(r_{i,j}/\sigma )
\qquad . 
\label{Pair}
\end{equation}
The essential assumptions are pairwise additivity and the fact that the pair
potential can be written as an energy parameter $\epsilon $ times a function
of the reduced distance $r/\sigma $. The Law of Corresponding States follows
when we assume that the pair potentials of all substances to which the law
applies are {\em conformal}. Interaction potentials are said to be {\it %
conformal} if their plots can be made to superimpose by
adjusting the values of $\epsilon $ and $\sigma $. With these assumptions,
the partition function is of the form: 
\begin{equation}
Q(N,V,T)=\left[ \frac{\sigma ^{3}}{\Lambda ^{3}}g\left( T^{*},\rho
^{*}\right) \right] \qquad ,  \label{Partition}
\end{equation}
where $g$ is the {\it same function} for all molecules and depends only on 
$T^{*}=k_{B}T/\epsilon $, the reduced temperature, and $\rho ^{*}=N/V\sigma
^{3}$ the reduced density\cite{McQuarry}. It then follows that many other
thermodynamic properties - in particular the pressure - are functions of 
$T^{*}$ and $\rho ^{*}$ only.

Unfortunately the interactions between real molecules are never truly
pairwise additive, nor are the pair potentials of different molecules
conformal. Even for inert gases the conformality of pair-potentials is only
fair. Moreover, the importance of three-body forces restricts the validity
of the assumption of pairwise additivity. While only a small family of
substances can be described by the original form of the Law of Corresponding
States, many fluids conform quite accurately to extended equations-of-state
that involve a third parameter. Thus, the compressibility factor $z$ can be
expressed as 
\begin{equation}
z=\beta P/\rho =f\left( T^{*},\rho ^{*},x\right) 
\qquad ,  
\label{CompFact}
\end{equation}
where $x$ is a third parameter that is usually related to some
characteristic feature of the phase diagram of a substance. At
first\cite{Hougen}  the critical compression factor $z_{c}$ was used
for $x$, but $ z_{c} $ is hard to determine with high accuracy, and a
better choice was sought.  The slope of the vapor pressure curve (at a
reduced temperature of $T_r=0.7$),  $\omega$, turned out to be a more
convenient choice for $x$ \cite{Riedel,Pitzer.2}. Various equations of
the form $z(T^{*},\rho ^{*},\omega)$ have been
presented\cite{Schreiber}, which agree well with the thermodynamic properties
for several classes of molecular fluids.\\

In this Communication we focus on the effect of changing the range of
attractive forces in suspensions of spherical colloids. As the range
of attraction varies independently of the hard-core radius $\sigma$,
the effective interactions are clearly not conformal. It is known that
the phase behavior of colloidal suspensions depends strongly on the
range of the attractive interactions. However, at present, there is
--to our knowledge-- no extended law of corresponding states that
allows us to make predictions about the phase behavior on the basis of
the effective pair potential alone. In fact, a wide variety of
non-conformal pair potentials have been used to describe the
interactions between colloids with short ranged attraction. It is our aim to
formulate an extended law of corresponding states that allows us to
compare such different pair potentials. In particular, we have
considered the square-well model\cite{Elliot}, attractive Yukawa potentials
\cite{Lomba,Hagen}, 2n-n Lennard-Jones type
potentials\cite{Vliegenthart.1}, the $\alpha $-Lennard-Jones
potential used in the description of protein-protein
interactions\cite{Meijer,tenWolde}, an effective potential reproducing
the depletion attractive forces in colloid-polymer
mixtures\cite{Meijer.1},  and more complex potentials, which include
a repulsive barrier, {\it i.e.} the effective two-body potential for
mixtures of nonadditive asymmetric hard spheres \cite{Dijkstra}.  At
this stage we limit our analysis to the phase behavior {\it around the
critical point}, but our findings should be generalized to densities
away from the critical region, in the spirit of the extended law of
corresponding states. \\

We proceed to calculate the
scaling parameters ($\epsilon ,\sigma $ and $x$), which stem from the 
knowledge of the inter-particle potential alone, without any need for
further experimental measurement.
An
obvious choice for the length scale $\sigma $ is the {\it effective} hard core
diameter.  Some care has to
be taken in the calculation of $\sigma ^{EFF}$ for continuous potentials
(such as the Lennard-Jones 2n-n). According to the Weeks-Chandler-Andersen
(WCA) method, we separate the potential into attractive $v_{att}$ and
repulsive $v_{rep}$ parts\cite{Andersen}, and calculate the ``equivalent''
hard-core diameter for the repulsive part of the potential using the 
expression suggested by Barker\cite{Barker}: 
\begin{equation}
\sigma ^{EFF}=\int_{0}^{\infty }dr\left[ 1-e^{-v_{rep}(r)/k_{B}T}\right]
\qquad .  
\label{SigEff}
\end{equation}
Two parameters are needed to properly describe the role of attractions: an
energy scale and a second quantity related to the range of attraction. At
low temperatures, the potential energy per particle in the crystalline phase
is given by the value of the pair-potential at the nearest-neighbor
separation, multiplied by the number of neighbors (and divided by two, to
correct for double counting). This is independent of the functional form of
the potential. This makes\ $v(r_{min})$, the depth of the potential well,
our natural choice for the energy scale $\epsilon $. The third parameter
that we use is the reduced second virial coefficient, {\it i.e.} the second virial
coefficient $B_{2}$ divided by the second virial coefficient of hard spheres
with a diameter $\sigma ^{EFF}$. The second virial coefficient $B_{2}$ can
be easily calculated once the functional form of the potential has been
specified: 
\begin{equation}
B_{2}=2\pi \int_{0}^{\infty }dr\quad r^{2}\left[ 1-e^{-v(r)/k_{B}T}\right] 
\qquad ,
\label{B2}
\end{equation}
and the reduced second virial coefficient $B_{2}^{\ast }$ is defined as 
\begin{equation}
B_{2}^{\ast }\equiv \frac{B_{2}}{2\pi (\sigma ^{EFF})^{3}/3} 
\qquad .
\label{B2*}
\end{equation}
Note that all three parameters ($\sigma ^{EFF},\epsilon $ and $B_{2}^{\ast }$%
) can be computed {\it only} based on the intermolecular potential $v(r)$. 
In this sense, our approach differs from those extended corresponding states laws
that use experimental data to define appropriate scaling parameters. This is
particularly useful for the description of colloidal systems where the
topology of the phase diagram changes as the range of the attraction is
decreased. For instance, it would not be feasible to use the properties at
the critical point as scaling parameters, as the critical point may be
experimentally inaccessible for sufficiently short-ranged attractions. Our
working hypothesis is that for a wide range of colloidal materials,
the compressibility factor $z$ is a function of only three parameters, {\em %
viz.} \ the reduced temperature $T^{\ast }=k_{B}T/\epsilon $, the reduced
density $\rho ^{\ast }=N/V(\sigma ^{EFF})^{3}$ and the reduced second-virial
coefficient $B_{2}^{\ast }$ 
\begin{equation}
z=f\left( T^{\ast },\rho ^{\ast },B_{2}^{\ast }\right) 
\qquad .
\label{ComFact.1}
\end{equation}
In the literature onhard particles with short-ranged attraction, it is
conventional to express the reduced second-virial coefficient in terms of a
parameter $\tau $ that is defined through the the following equation
\cite{Baxter}:
\begin{equation}
B_{2}^{\ast }\equiv 1-\frac{1}{4\tau } 
\qquad .
\label{TauDef}
\end{equation}
$\tau $ is a measure for the temperature - low (high) $T$ corresponds to low
(high) $\tau $. However, $\tau $ is not a linear function of $T$.

%%%%%%%%%%%%%%%%%%%%%%%%%%%%%%%%%%%%%%%%%%%%%%%%%%%%%%%%%%%%%%
%                                                            %  
%  Figure 1                                                 
%                                                            %
%%%%%%%%%%%%%%%%%%%%%%%%%%%%%%%%%%%%%%%%%%%%%%%%%%%%%%%%%%%%%%
\begin{figure}
\begin{minipage}{6.7cm}
	\makebox[6.7cm][l]{$\tau$}
    	\epsfxsize 6.0 cm
    	\rotatebox{-90}{\epsfbox{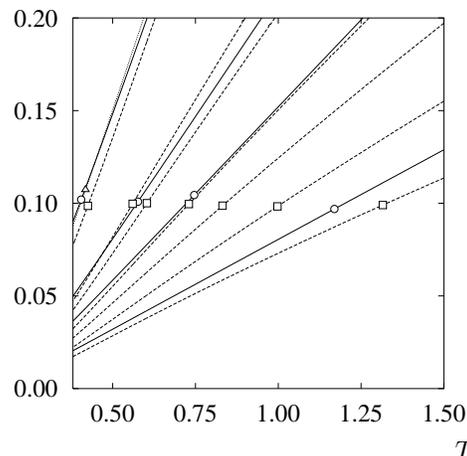}}
	\makebox[6.7cm][r]{$T^*$}
\end{minipage}
\caption{Stickyness parameter $\tau$ plotted versus the reduced temperature
$T^*=k_{B}T/\epsilon$ for different potentials. 
Circles: attractive Yukawa, 
%(\mbox{\raisebox{-.5ex}{\includegraphics[scale=0.1]{op_cir_so.ps}}}) 
attractive Yukawa, 
Squares: 
%(\mbox{\raisebox{-.5ex}{\includegraphics[scale=0.1]{op_squ_da.ps}}})
Lennard Jones 2n-2, 
Triangles: 
%(\mbox{\raisebox{-.5ex}{\includegraphics[scale=0.1]{op_tri_do.ps}}})
$\alpha$-Lennard Jones}
\label{tau-t}
\end{figure}

In Figure (\ref{tau-t}) we have plotted the stickyness parameter
$\tau$ as a function of the reduced temperature for some of the cases
listed in Table  \ref{CriticalPoints}. In the
temperature range studied, the stickyness parameter increases  almost
linearly with the reduced temperature\cite{Comment.1}. The figure shows another
important feature: if the $\tau-T^*$ curves for two different
potentials are close at any particular temperature, they tend to  be
close for {\it all} temperatures studied. Such behavior is an
indication  that the present scheme to compare non-conformal
potentials is reasonable.  
As can be seen from the figure
(and from Table \ref{CriticalPoints}), the value of $\tau $ -- and therefore 
that of
the reduced second virial coefficient -- at the critical point is remarkably
constant (around $\tau\approx 0.1$). This fact had been noted earlier by 
Vliegenthart 
and Lekkerkerker%
\cite{Vliegenthart}. In fact, $\tau $ hardly varies between the limit of
extremely narrow attractive wells (Baxter's adhesive hard-sphere model 
\cite{Baxter}) and the (van der Waals) limit of infinitely long-ranged 
attractive
wells. Also in models that are a mixture
of the Baxter and van der Waals model the value of $\tau $ at the critical
point varies only slightly\cite{Noro}.

We mentioned above that the reduced second virial coefficient is a measure
for the range of the attractive part of the potential. To make this
statement more precise, we have to specify what we mean by the ``range'' of
a potential. Here, we take the following route: there is one system for
which the range of the attractive potential is defined unambiguously,
namely hard spheres with a square-well attraction 
\begin{equation}
u(r)=\left\{ {\ \matrix {\ \infty \hfill \cr -\epsilon \hfill \cr 0 \hfill %
\cr }}\right. \quad \matrix {\ {r\le \sigma } \hfill \cr {\sigma <r \le \lambda
\sigma} \hfill \cr {\lambda \sigma <r} \hfill \cr }
\qquad .
\end{equation}

A logical choice for a dimensionless measure for the range of the attractive
part of the potential is $R\equiv \lambda -1$. In the spirit of our extended 
corresponding
states approach, we now {\em define} \ the range of an arbitrary attractive
potential to be equal to the range of that square-well potential that yields
the same reduced second virial coefficient at the same reduced temperature.
The reduced second virial coefficient of a square-well potential is given by 
\begin{equation}
B_{2}^{*}=1 + \left[ \lambda^{3}-1\right] \left( 1-e^{1/T^{*}}\right) 
\qquad ,
\end{equation}
and hence
\begin{equation}
\tau =\frac{1}{4\left[ \lambda^{3}-1\right] \left( e^{1/T^{*}}-1\right) }
\qquad .
\label{tauB2}
\end{equation}
Using this mapping onto the square-well system, we have computed the effective
range of the attractive part of the potential for a number of different
potential functions that have been used to describe colloidal suspensions or
globular protein solutions. In general, the range of attraction is still
temperature dependent. In Table \ref{CriticalPoints}, we have collected the
values of $R$ at the temperature corresponding to the liquid-vapor 
critical
point. In the same table, we also give the value for the ``stickyness''
parameter $\tau $ at the critical temperature.

%%%%%%%%%%%%%%%%%%%%%%%%%%%%%%%%%%%%%%%%%%%%%%%%%%%%%%%%%%%%%%
%                                                            %  
%  Figure  2                                                 %
%                                                            %
%%%%%%%%%%%%%%%%%%%%%%%%%%%%%%%%%%%%%%%%%%%%%%%%%%%%%%%%%%%%%%
\begin{figure}[tbp]
\begin{minipage}{6.7cm}
	\makebox[6.7cm][l]{$T^*_c$}
    	\epsfxsize 6.0 cm
    	\rotatebox{-90}{\epsfbox{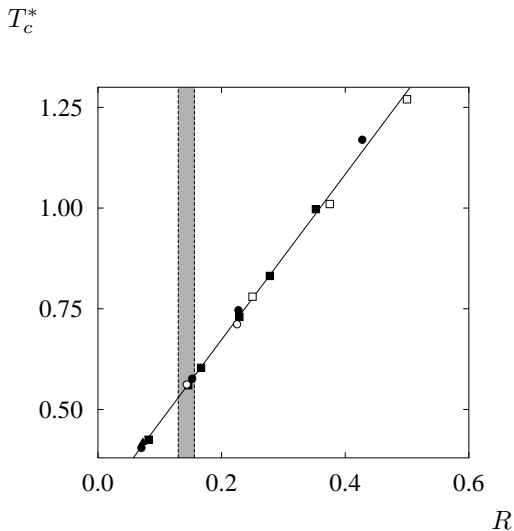}}
	\makebox[6.7cm][r]{$R$}
\end{minipage}
\caption{The reduced temperature at the liquid-gas critical point 
$T^*_c=k_{B}T_c/\epsilon $  plotted versus the range of attractions  
$R=\lambda -1$ of the
equivalent square well system. As the range becomes shorter than the
threshold value $\approx 0.14$,  the liquid-gas transition becomes
metastable. 
Filled Circles: 
%(\mbox{\raisebox{-.5ex}{\includegraphics[scale=0.1]{cl_cir.ps}}}) 
attractive Yukawa, 
Filled Squares: 
%(\mbox{\raisebox{-.5ex}{\includegraphics[scale=0.1]{cl_squ.ps}}}) 
Lennard Jones 2n-n, 
Filled Triangles: 
%(\mbox{\raisebox{-.5ex}{\includegraphics[scale=0.1]{cl_tri.ps}}}) 
$\alpha$-Lennard Jones, 
Open Squares: 
%(\mbox{\raisebox{-.5ex}{\includegraphics[scale=0.1]{op_squ.ps}}}) 
Square Well, 
Open Circles: 
%(\mbox{\raisebox{-.5ex}{\includegraphics[scale=0.1]{op_cir.ps}}}) 
effective Colloid-Colloid interaction}
\label{t-R}
\end{figure}

In Figure  (\ref{t-R}), we show the relation between
$T_{c}^{*}$, the reduced critical temperature and $R$, the range of
the attractive potential. In the temperature range studied, the
relation between $R$ and  $T_{c}^{*}$ is surprisingly linear --
although, again, we know that this linear relation cannot hold for
values of $R$ very close to zero -- and obeys the simple relation:
\begin{equation}
T_{c}^{*}\approx 0.26 + 2.1 R 
\qquad .
\end{equation}
The range of the attractive part of the potential determines whether a
given system can exhibit a stable liquid-vapor transition or whether this
transition is preempted by freezing. The disappearance of the
liquid-vapor transition in systems with short-ranged attraction was
first noted in theoretical work by Gast, Russel and Hall\cite{Gast}.
This work has subsequently been placed on a firmer theoretical footing
by Lekkerkerker et al\cite{WarrenPoonLekkerkerker}. Evidence for the
disappearing of the liquid-vapor critical point comes from both 
simulation\cite{Meijer,Hagen} and
experiment\cite{PoonIllet}.  All authors agree that the liquid-vapor
transition disappears for sufficiently short-ranged attraction.
However, estimates differ for the value of $R$ where this change in
the phase diagram takes place. Estimates for $R$ vary from $0.1$ to
almost $0.4$.  Part of the reason why the different estimates for the
critical value of $R$ appear inconsistent is that the various authors
have studied systems with non-conformal interaction potentials and,
more importantly, have used different definitions for the range. 
The advantage of the present approach is that we have a unique
way to define the range of the attractive potential for widely
different interaction potentials.  
When we consider the
available data for the  2n-n Lennard-Jones potentials, the $\alpha $%
-Lennard-Jones potential, and the attractive Yukawa system, we find
that in every case the boundary between stable and meta-stable
liquid-vapor transitions is located within a narrow band between
$R=0.13$ and  $R=0.15$.  In Figure (\ref{t-R}) critical points plotted
to the right of the vertical $R\approx 0.14$ band refer to a  {\it
stable} transition, while points to the left are
metastable. 
To date no simulation has computed the
threshold value for square-well particles. A rough
estimate of $R\approx 0.25$ has been calculated using a simple van der Waals
model for both the fluid and the solid phase\cite{Tejero} and from a 
simple cell model with some phenomenological character\cite{Benedek}. 
For $R=0.85$, theoretical estimates suggest that in this case the
critical point is stable\cite{Lincoln}.

The predictive power of our approach based on  
pair potentials is expected to break down when three-body
interactions become important.  We have also tested our theory for
more complex pair potentials, which included a repulsive
barrier\cite{Dijkstra}. Here too we have found deviations from
extended corresponding states behavior: in several cases the calulated
$\tau$ parameter,  at the critical point, lies much below the constant
value of $0.10$, and the  mapping onto the equivalent square-well
system yields unphysically small attraction ranges $R$. The repulsive
barrier alters the effective size of the particle, especially at low
temperatures, but our WCA decomposition of the effective potential limits the
repulsive contribution to the hard body term.

In summary, we have formulated a simple extended corresponding states
principle that allows us to make predictions about the topology of the phase
diagram of suspensions of spherical colloids with variable range attraction.
The scaling parameter $\sigma $, $\epsilon $ and $B_{2}^{*}$ can all be
derived directly from knowledge of the pair-potential. Moreover, this
procedure allows us to give an unambiguous definition for the range of
the attractive part of the potential. By analysing a number of
simulation data for different model systems, we find that the liquid-vapor
transition becomes metastable with respect to the freezing transition when
the range of the attraction becomes less than approximately $0.14$.

$\star\star\star$\\
The work of the FOM Institute is part of the research program of ``Stichting
Fundamenteel Onderzoek der Materie'' (FOM) and is supported by NWO. We
gratefully acknowledge discussions with Fernando Fernandez. We also 
thank N.Kern and H.N.W.Lekkerkerker for critical reading of the manuscript 
and for useful suggestions.
MGN acknowledges financial support from EU contract ERBFMBICT982949.

%%%%%%%%%%%%%%%%%%%%%%%%%%%%%%%%%%%%%%%%%%%%%%%%%%%%%%%%%%%%%%
%                                                            % 
%  TABLE I                                                   %
%                                                            %
%%%%%%%%%%%%%%%%%%%%%%%%%%%%%%%%%%%%%%%%%%%%%%%%%%%%%%%%%%%%%%
\begin{table}[tbp]
\begin{tabular}{lccc}
& $T^*_c$ & $\tau$ & R \\ \hline
Square Well\cite{Elliot} &  &  &  \\ 
& 2.61 & 0.0765 & 1.000 \\ 
& 1.79 & 0.0766 & 0.750 \\ 
& 1.27 & 0.0942 & 0.500 \\ 
& 1.01 & 0.0924 & 0.375 \\ 
& 0.78 & 0.1007 & 0.250 \\ \hline
Yukawa\cite{Lomba,Hagen} &  &  &  \\ 
& 1.170 & 0.0969 & 0.427 \\ 
& 0.715 & 0.1044 & 0.227 \\ 
& 0.576 & 0.1009 & 0.153 \\ 
& 0.412 & 0.1020 & 0.070 \\ \hline
2n-n\cite{Vliegenthart.1} &  &  &  \\ 
& 1.316 & 0.0990 & 0.476 \\ 
& 0.997 & 0.0983 & 0.353 \\
& 0.831 & 0.0987 & 0.278 \\ 
& 0.730 & 0.0996 & 0.229 \\
& 0.603 & 0.1001 & 0.167 \\ 
& 0.560 & 0.0997 & 0.146 \\ 
& 0.425 & 0.0986 & 0.082 \\ \hline
$\alpha$-LJ\cite{Meijer} &  &  &  \\ 
& 0.418 & 0.1073 & 0.073 \\ \hline
Colloid\cite{Meijer.1} &  &  &  \\ 
& 0.712 & 0.0970 & 0.225 \\ 
& 0.562 & 0.1023 & 0.144 \\ \hline
Asymm. hard-spheres\cite{Dijkstra} &  &  &  \\ 
& 0.186 & 0.0744 & 0.005 \\
& 0.173 & 0.0758 & 0.003 \\
& 0.164 & 0.0788 & 0.002
\end{tabular}
\caption{Values of $T^*$, $\tau$ and the range of the equivalent square well
system $R$ for different potentials, and for different ranges, 
calculated at
the liquid-gas critical point}
\label{CriticalPoints}
\end{table}


\begin{references}
\bibitem{Comment.1}
However, as can be seen
from Eqn. (\ref{tauB2}), the linear relation should break down  for
sufficiently low reduced temperatures.
\end{references}
\end{document}